\documentclass[%
reprint,
 amsmath,amssymb,
 aps,
 pra,
]{revtex4-1}

\usepackage{graphicx}% Include figure files
\usepackage{epsfig}
\usepackage{color}% Include figure files
\usepackage{dcolumn}% Align table columns on decimal point
\usepackage{bm}% bold math
\usepackage{hyperref}% add hypertext capabilities

\begin{document}

\title{How to correctly quantify neuronal phase-response curves from noisy recordings} 

\author{Janina Hesse}
  \email{janina.hesse@bccn-berlin.de}
\author{Susanne Schreiber}
\affiliation{%
Institute for Theoretical Biology, 
Department of Biology,
Humboldt-Universit\"at zu Berlin,
\\
Philippstr. 13, Haus 4, 10115 Berlin\\
Bernstein Center for Computational Neuroscience Berlin
 }

\date{\today}% It is always \today, today,

% \PACS{PACS code1 \and PACS code2 \and more}
% \subclass{MSC code1 \and MSC code2 \and more}

\begin{abstract}
At the level of individual neurons, various coding properties can be inferred from the input-output relationship of a cell. For small inputs, this relation is captured by the phase-response curve (PRC), which measures the effect of a small perturbation on the timing of the subsequent spike. Experimentally, however, an accurate experimental estimation of PRCs is challenging. Despite elaborate measurement efforts, experimental PRC estimates often cannot be related to those from modeling studies. In particular, experimental PRCs rarely resemble the generic PRC expected close to spike initiation, which is indicative of the underlying spike-onset bifurcation. Here, we show for conductance-based model neurons that the correspondence between theoretical and measured phase-response curve is lost when the stimuli used for the estimation are too large. In this case, the derived phase-response curve is distorted beyond recognition and takes on a generic shape that reflects the measurement protocol, but not the real neuronal dynamics. We discuss how to identify appropriate stimulus strengths for perturbation and noise-stimulation methods, which permit to estimate PRCs that reliably reflect the spike-onset bifurcation -- a task that is particularly difficult if a lower bound for the stimulus amplitude is dictated by prominent intrinsic neuronal noise.
\end{abstract}

\maketitle

\section{Introduction}
\label{intro}

Neuronal dynamics are commonly studied based on the
response of a neuron to specific stimuli. The spiking in response to post-synaptic currents or temporally structured inputs, for example, often serves as a first indication for the neuronal code.
Constraining the response to weak stimuli, a neuron's spiking dynamics can be
captured by the so-called \emph{phase-response curve} (PRC). The PRC relates the timing of a short
stimulus pulse to the consequential advance or delay of the next spike. As a well-defined theoretical quantity, the PRC is also experimentally accessible. PRCs are valuable to decipher single-cell dynamics, such as the neuronal response to a particular input shape, e.g. synaptic inputs \citep{reyes_effects_1993,netoff_synchronization_2005} or noise \citep{ermentrout_reliability_2008}, as well as the locking time to a time-dependent stimulus \citep{kuramoto_chemical_1984,achuthan_synaptic_2011}. Beyond single cells, PRCs are often used to predict how neurons behave when weakly connected, informing about network dynamics and, in particular, their synchronization state \citep{van_vreeswijk_when_1994,hansel_synchrony_1995,ermentrout_type_1996,galan_efficient_2005,teramae_temporal_2008,smeal_phase-response_2010}. Beyond the neurosciences, the PRC is used for various other biological oscillations such as cardiac pacemakers or circadian rhythms \citep{guevara_phase_1981,minors_human_1991}.

As we show in the following, suboptimal measurement protocols yield PRCs that do not accurately reflect neuronal dynamics. In order to estimate PRCs experimentally, neuronal spiking is perturbed by an experimentally known stimulus. The amplitude range of this stimulus is bounded from below and from above: On one hand, the experimental stimulus has to be chosen large enough to overcome intrinsic noise, which perturbs spiking in addition to the stimulus \citep{manwani_detecting_1999}. On the other hand, the stimulus has to be weak enough such that the PRC is a valid description of the dynamics: Stimulus amplitudes have to be sufficiently small to allow the dynamics to relax back to
the limit cycle within one period, because the PRC assumes independence of subsequent spikes. As a rule of thumb, the larger the intrinsic noise, the more restricted the range of acceptable stimulation amplitudes.

Relatively strong intrinsic noise is, for example, typical for cortical and hippocampal neurons, which often show notable spike jitter even for constant step current stimuli. We show that the consequential variability of the phase response observed experimentally can impair PRC measurements \citep{wang_hippocampal_2013,stiefel_neurons_2016}. This was not observed previously because the methods for quantification of PRCs were typically illustrated with neurons showing only low levels of intrinsic noise, and thus stable firing rates (for a review see \citet{torben-Nielsen_comparison_2010}).

Here, we address the identification of spike-onset bifurcations from PRC measurements. Spiking in neurons can be classified into four basic mechanisms of spike initiation \citep{izhikevich_dynamical_2007}. The corresponding spike-onset bifurcations give rise to distinguishable, generic PRCs that reflect neuronal dynamics at spike onset \citep{ermentrout_type_1996,izhikevich_neural_2000}.
We show that the mapping between PRC and spike-onset
bifurcation is lost when the measurement stimulus is not chosen appropriately.
In this case, the PRC shape becomes
independent of the neuron's dynamics, and only reflects the measurement protocol. 

In the following, we compare the measurement accuracy of three different
PRC methods faced with strong intrinsic noise. While \citet{torben-Nielsen_comparison_2010} have previously considered different methods under relatively benign
conditions, we are interested in
extreme conditions with large stimuli or large intrinsic noise levels. 
Exploring the border where a PRC estimation becomes invalid allows us to
delineate the range of experimental conditions under which PRC
measurements can be expected to actually yield information on neuronal dynamics.

\section{Methods}\label{methods}

\subsection{Models}

PRCs were measured for conductance-based neuron models with three
different spike onset bifurcations, for details see Appendix. Of the four possible limit cycle bifurcations for spike initiation \citep{izhikevich_dynamical_2007}, we have chosen those that result in biologically realistic all-or-none spikes with finite voltage amplitude at spike onset: the subcritical Hopf bifurcation, the saddle-node on invariant cycle (SNIC) bifurcation and the saddle-homoclinic orbit (HOM) bifurcation. The level of the intrinsic noise was set by the standard deviation of an additive zero-mean white-noise current, see Appendix.

\subsection{Phase-response curve}

For a regular spiking conductance-based neuron model with baseline period $T$ (firing rate $f = 1/T$), weakly perturbed spiking can be described by an input-output equivalent phase oscillator, 
\begin{equation}
\dot{\varphi} = 1 + Z(t) s(t), \label{eq:phase}
\end{equation}
with $Z$ as PRC, and a time-dependent stimulus $s(t)$. In response to the stimulus $s(t)$, the spike at $t=t_{i+1}$ following the spike $t=t_i$ is advanced or delayed according to the phase deviation 
\begin{equation}
\Delta\varphi = \int_{t=t_i}^{t_i+T} Z(t) s(t) \textrm{d}t. \label{eq:phaseAdvance}
\end{equation}

\subsection{Theoretical estimation of the PRC}\label{backward-integration-of-the-adjoint-equation}

For mathematical neuron models, the PRC can be gained by numerical
integration, for a review see \citet{govaerts_computation_2006}. For a conductance based neuron model with equation
$\dot{x} = F(x)$, the linearized dynamics on the limit cycle are given
as $\dot{x} = J x$ with the Jacobian
$J = \frac{\partial F}{\partial x}$. The corresponding adjoint
equation is $\dot{y} = -J^{\mathrm{T}} y$. As these dynamics are
unstable, backward integration in time along the limit cycle leads to a
stable solution $y$, whose first entry corresponds to the voltage PRC \citep{ermentrout_multiple_1991,ermentrout_type_1996}.
We use this method to derive the ``true'' PRC for the models. The comparison of theoretical PRC and PRCs estimated with experimentally-inspired methods allows us to suggest the appropriate 
input strength, as well as a reasonable noise level,
both of which can be employed in biological experiments to prevent 
false PRC measures as mentioned in the introduction.

\subsection{Experimental methods of PRC measurements
}\label{experimentally-inspired-measurements-of-the-prc}

We consider three different methods for PRC estimation. The estimation analyzes spike timing in response to a known stimulus. For the first method, neuronal spiking is perturbed by a sequence of short current pulses, the other two method use a noise stimulus. In addition, the neuron is stimulated with a DC current that ensures repetitive spiking; we choose a baseline firing rate of 10Hz (i.e., with period $T$ = 100ms). The simulation duration is sufficient to record about 500 perturbed spikes, for details see Appendix.

For
two consecutive spikes at time $t=t_i$ and $t = t_{i+1}$, the actual
interspike interval duration is $\textrm{ISI}_i = t_{i+1} - t_i$. For the phase response, we normalize the time during the unperturbed period $T$ to a phase variable $\varphi$ that ranges from zero to one. For a perturbed interspike interval, the change in duration is captured by the \emph{phase deviation}, $\Delta\varphi_i = 1 - (t_{i+1} - t_i)/T$, which is
positive for spike advances, and negative for phase delays. 

For plotting purposes, the PRCs are commonly fitted either with polynomial functions, e.g., \citet{netoff_beyond_2005}, or with Fourier series consisting of a low number of Fourier coefficient, e.g., \citet{galan_efficient_2005}. As we are here also interested in PRCs with steep components (such as the PRC of the HOM model in Figure~\ref{fig:PRCs}a), we choose a Fourier series of order five (i.e., with 11 Fourier components) to fit our PRC estimates, 
\begin{equation}
Z(\varphi) = a_0 + \sum_{j = 1}^5 \left(a_j \cos(2\pi j \varphi) + b_j \sin(2\pi j \varphi)\right), \label{eq:fourierSum}
\end{equation}
with Fourier coefficients ${a_0,a_1,...a_5, b_1, ..., b_5}$. Using the same number of Fourier components for all PRC estimates facilitates the comparison of different methods.

\begin{figure*}[h]
\hfill
\begin{center}
\includegraphics[width=0.75\textwidth]{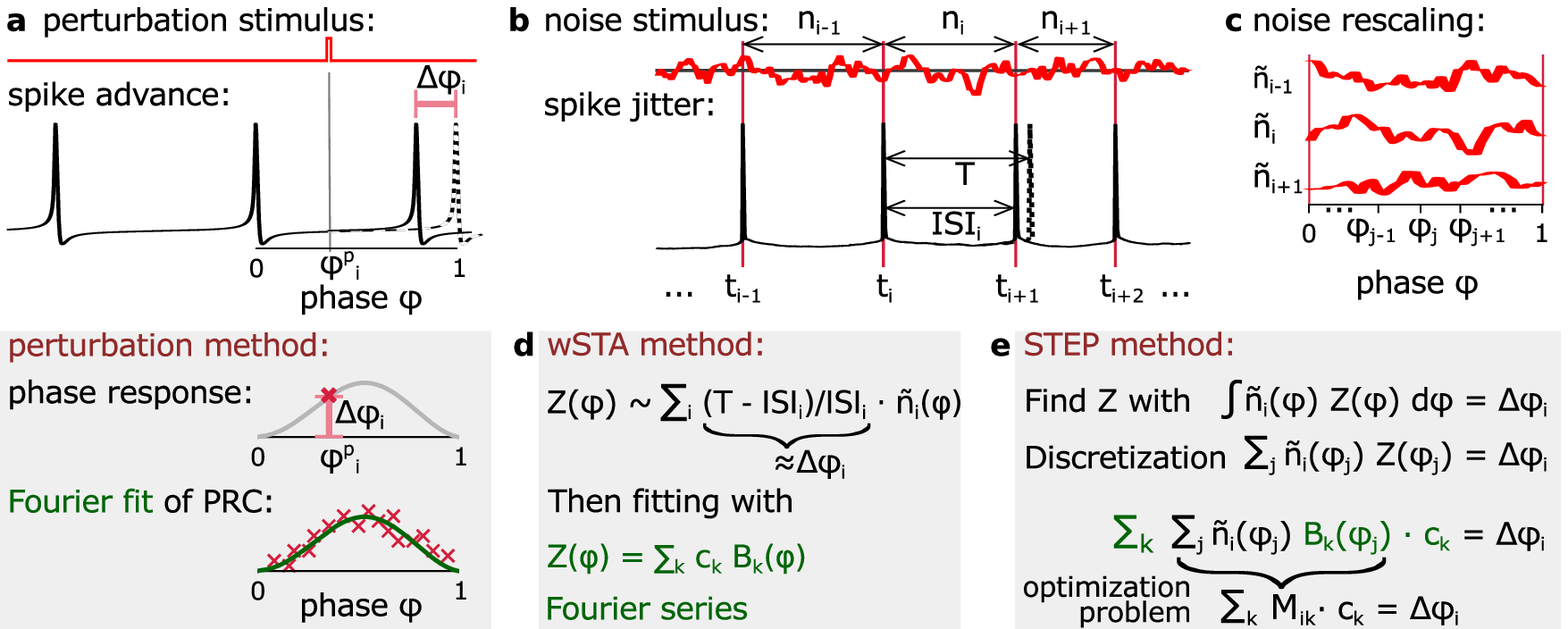}
\end{center}
\caption{Experimental methods for PRC measurements. a: Individual time-point perturbation. The phase deviation $\Delta \varphi$, resulting from stimulation with a short current pulse, is plotted against the timing of the current pulse, and fitted by a curve. b: Time-continuous noise
stimulation. Spikes are jittered due to a continuous noise-stimulation. c: Noise is rescaled to the phase variable. d: The wSTA method estimates the PRC via a weighted spike-triggered average. e: The STEP method relies on an error minimization of the input-and-PRC-based prediction of spike jitter.}
\label{fig:PRCsMeasure}
\end{figure*}

\subsubsection{PRC estimation based on individual-timepoint
perturbation}\label{individual-timepoint-perturbation}

For the PRC estimation based on individual-timepoint perturbations, the repetitively spiking neuron is stimulated with delta current
pulses of a specific amplitude. The perturbation
method was chosen for most experimentally measured PRCs \citep{reyes_two_1993,reyes_effects_1993,goldberg_response_2007,tsubo_layer_2007,wang_hippocampal_2013}, as it is
intuitive and easily analyzed.
Here, the interval between two successive pulses is chosen randomly between
150ms and 250ms, such that most perturbations occur as single events in
every second interspike interval. This ensures the independence of
subsequent perturbations, allowing the
dynamics a whole period to relax.

The analysis evaluates interspike intervals in which a current
pulse occurs. The timing of the perturbation is related
to the resulting phase deviation, compare Figure~\ref{fig:PRCsMeasure}a. Given a current pulse at $t=t_{p}$ between two consecutive spikes at
$t=t_i$ and $t=t_{i+1}$, the PRC relates the phase deviation, $\Delta\varphi_i = 1 - (t_{i+1} - t_i)/T$, to the phase at which the
current pulse occurred, $\varphi^{p}_i = (t_{p} - t_i)/T$ \citep{rinzel_analysis_1989}.
An estimate of the PRC in units of
phase deviation is gained by plotting the phase deviation against the
perturbation phase. More precisely, each point of this graph corresponds
to a stochastic drawing from the underlying PRC. In order to get the PRC in
response to current perturbations, the PRC in units of phase deviation
is divided by the amplitude of the perturbation stimulus.

For neurons with intrinsic noise, even weak noise perturbs the raw data so strongly that the PRC shape can often not be identified by naked eye. To fit a curve to the raw PRC data, we here use Gal{\'a}n's method \citep{galan_efficient_2005,netoff_experimentally_2011}, for which the Fourier components of Eq.~(\ref{eq:fourierSum}) are given for a set of $N$ perturbed spikes as
\begin{eqnarray*}
a_0 = \frac{1}{N} \sum_{i=1}^N \Delta\varphi_i, 
\quad a_j = \frac{1}{N} \sum_{i=1}^N \Delta\varphi_i \cos(2\pi j \varphi^{p}_i),\\
\quad b_j = \frac{1}{N} \sum_{i=1}^N \Delta\varphi_i \sin(2\pi j \varphi^{p}_i).
\end{eqnarray*}

%(p.117)

\subsubsection{PRC estimation based on time-continuous noise
stimulation}\label{time-continuous-noise-stimulation}

For the PRC estimation based on noise-stimulation, the repetitively spiking neuron is stimulated with a zero-mean noise stimulus. We use a colored noise current with time resolution of $dt=10\mu\textrm{s}$, resulting from filtering a
white noise signal with a cut-off frequency of 1000Hz.

For each spike, the
recorded phase deviation $\Delta \varphi_i$ is related to the noise snippet $n_i$ that
corresponds to the preceding interspike interval, see Figure~\ref{fig:PRCsMeasure}b. The duration of the
noise snippet $n_i(t)$ is rescaled to the phase variable, resulting in a phase-dependent noise
snippet $\tilde{n}_i(\varphi)$, see Figure~\ref{fig:PRCsMeasure}c.

To estimate PRCs from the noise-stimulated spike trains, we use the \emph{weighted Spike-Triggered
Average} (wSTA) introduced by \citet{ota_weighted_2009} and the
\emph{STandardized Error Prediction} (STEP)  method introduced by
\citet{torben-Nielsen_novel_2010}. 

The wSTA method sums over the phase-dependent noise
snippets, while weighting each noise snippet by a variant of the phase
deviation $\bar{\Delta\varphi} = (T/(t_{i+1} - t_{i}) - 1)$, see Figure~\ref{fig:PRCsMeasure}d. For the correct amplitude scaling, the result is then divided by the variance of the noise input.

The STEP method optimizes the PRC shape to predict the phase deviation
caused by the noise input since the previous spike. The idea takes advantage of Eq.~(\ref{eq:phaseAdvance}), which predicts the phase deviation resulting from the stimulus $s(t) = n_i(t)$. Both the PRC and the phase-dependent noise snippets are discretized by a temporal binning with 200 phase
bins. This allows to replace the integration in Eq.~(\ref{eq:phaseAdvance}) by a simple sum of the PRC and the phase-dependent snippet $\tilde{n}_i$  (Figure~\ref{fig:PRCsMeasure}e). To create an optimization matrix $\mathbf{{M}}$, the
binned versions of noise and base functions are multiplied (Figure~\ref{fig:PRCsMeasure}e). Linear least square optimization
(here we used the python function \texttt{numpy.linalg.lstsq()}) allows
to find the Fourier coefficients that, when multiplied by the matrix $\mathbf{{M}}$, best recover the phase deviations extracted from the raw data of spike times.

An advantage of the perturbation method and the wSTA
method compared to the STEP method is that the PRC estimates can be
implemented as an ongoing process that continuously allows to add new
data as it is available. In contrast, the STEP method, and similar
methods that rely on spike prediction error minimization, require a set
of noise/spike-timing pairs of a fixed size, and, at least in current
implementations, the optimization has to be redone when new data is
collected.

\section{Results}\label{results}

In order to measure PRCs experimentally from repetitively firing neurons, the spike times are perturbed by an additional stimulus consisting of either a noise current or short pulse-like perturbations. Under the assumption that the stimulus has to be weak, intrinsic noise in real neurons makes
the appropriate scaling of the stimulus amplitude a non-trivial problem
\citep{netoff_experimentally_2011}. The stimulus has to be large enough to stand out against the intrinsic noise, yet also small enough to prevent the instantaneous induction of spikes.

While most previous studies focus on the PRC shape, and neglect the PRC
amplitude, we here also evaluate the PRC amplitude. The PRC amplitude is
essential for a quantitative comparison of measurements, as it scales quantities 
derived from the PRC such as the synchronization range or the transition time until
locking is established.

\subsection{PRC estimates with perturbations or noise stimuli}

\begin{figure*}[h]
\hfill
\begin{center}
\includegraphics[width=0.75\textwidth]{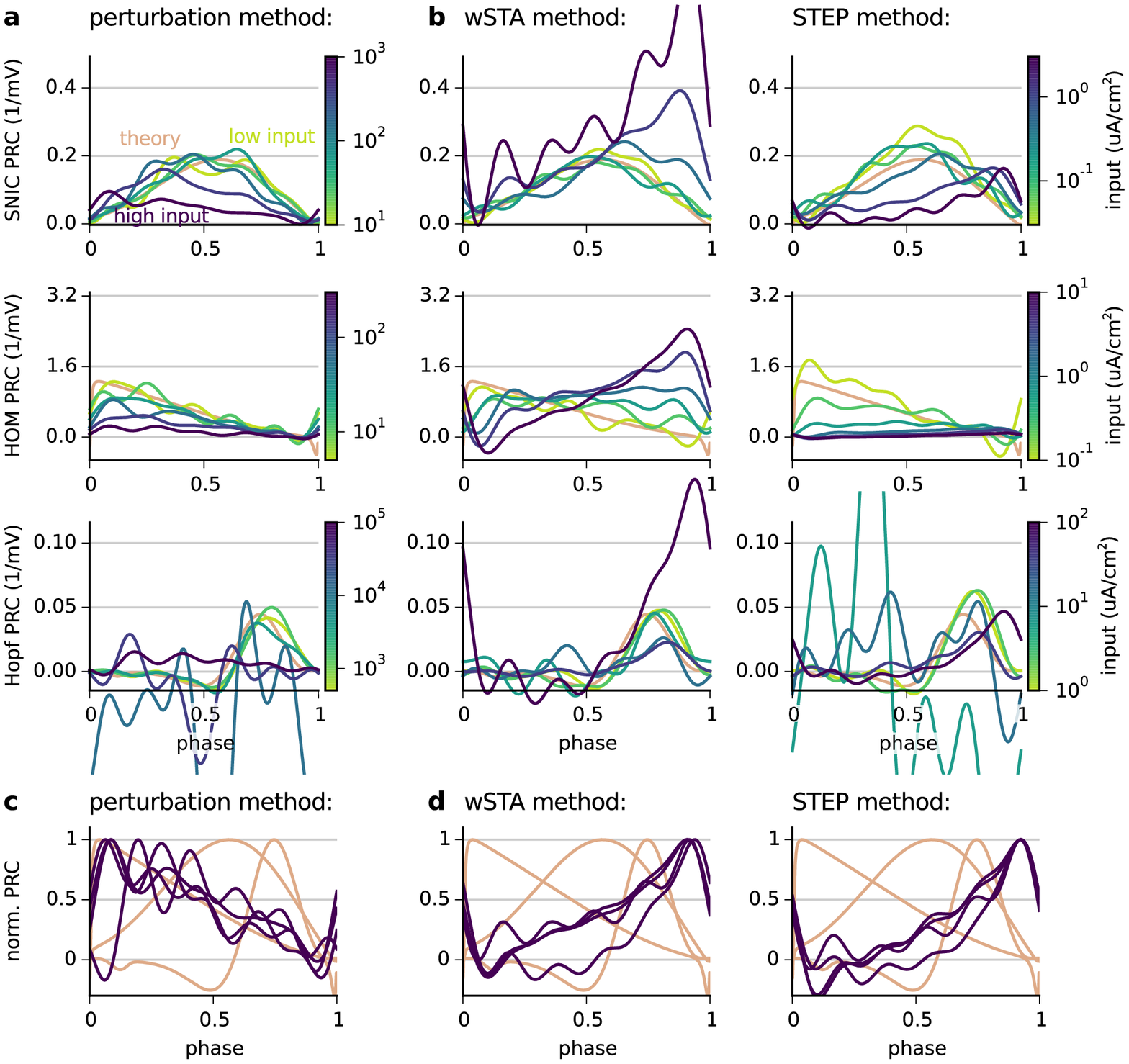}  % was 20180828_PRCs.pdf
\end{center}
\caption{Comparing PRC measurements for three spike generation mechanism (input strength is color-coded from green to violet). The smooth theoretical PRC is depicted in brown. a: PRCs measured with the perturbation method. b: PRCs measured with wSTA and STEP method. c-d:
Comparison of different dynamical classes at highest stimulus amplitude from
(a-b).}
\label{fig:PRCs}
\end{figure*}

Figure~\ref{fig:PRCs}a-b shows PRC estimates for models without intrinsic noise with three different spike generation mechanisms (top to bottom), using three different methods (left to right: perturbation method, wSTA method and STEP method), see Methods for details. For comparison, each small panel depicts the theoretical PRC (brown) and PRCs estimated from spiking in response to different stimulus strengths (color-coded from green to violet). While all three methods estimate the shape of the PRC with similar quality, the STEP method has a tendency to overestimate the amplitude of the PRC, which is not observed for the perturbation method and the wSTA method.

For the Hopf model, the PRC derived with the wSTA method seems reasonable. In contrast, the perturbation method and the STEP method both lead to strongly 
wiggling PRCs for intermediate stimulus amplitudes, see Figure~\ref{fig:PRCs}a-b. 
The wiggling PRCs occur for stimulus amplitudes for which the
model becomes extremely sensitive to inputs (the CV shows a maximum in
response to these noise strength).
As a side-note, this behavior was also observed for another model with
subcritical Hopf bifurcation at spike onset, the original Hodgkin-Huxley
model \citep{hodgkin_quantitative_1952}.

\subsection{Stimulus amplitude
dependence}\label{stimulus-amplitude-dependence}

The range of stimulus amplitudes in Figure~\ref{fig:PRCs}a-b is chosen to illustrate the transition from good to bad PRC estimates. Because the models lack intrinsic noise, the theoretical curve (brown) is best fitted by the PRC with the lowest input strength (green), and reasonable fits are derived for a range of small stimulus levels.
The correspondence between theoretical and
estimated PRC is lost for intermediate stimulus levels, and the estimated PRC approaches a characteristic shape for large stimulus
levels. Figure~\ref{fig:PRCs}c-d summarizes the PRC estimates
corresponding to the largest stimulus strength (the violet curves from Figure~\ref{fig:PRCs}a-b). For all methods, these PRCs have a characteristic shape that is hardly
distinguishable for different spike generation mechanisms, and thus the PRC is not informative about neuronal dynamics. With such a large input, the
drive is too strong to estimate meaningful PRCs, and we call this regime in the following the \emph{overdriven} regime.

\subsection{Shape of overdriven PRCs}\label{shape-of-overdriven-prcs}

The stereotypical PRC shape in the overdriven regime (Figure~\ref{fig:PRCs}c-d) shows a linear relation for intermediate phases, with a steep connection around phase zero that is enforced by the periodicity of
the Fourier-series fit. The linear decrease/increase observed for the perturbation and noise-stimulation methods, respectively, results from instantaneous spikes in response to large perturbations in the stimulus. 

\begin{figure*}[h]
\hfill
\begin{center}
\includegraphics[width=0.5\textwidth]{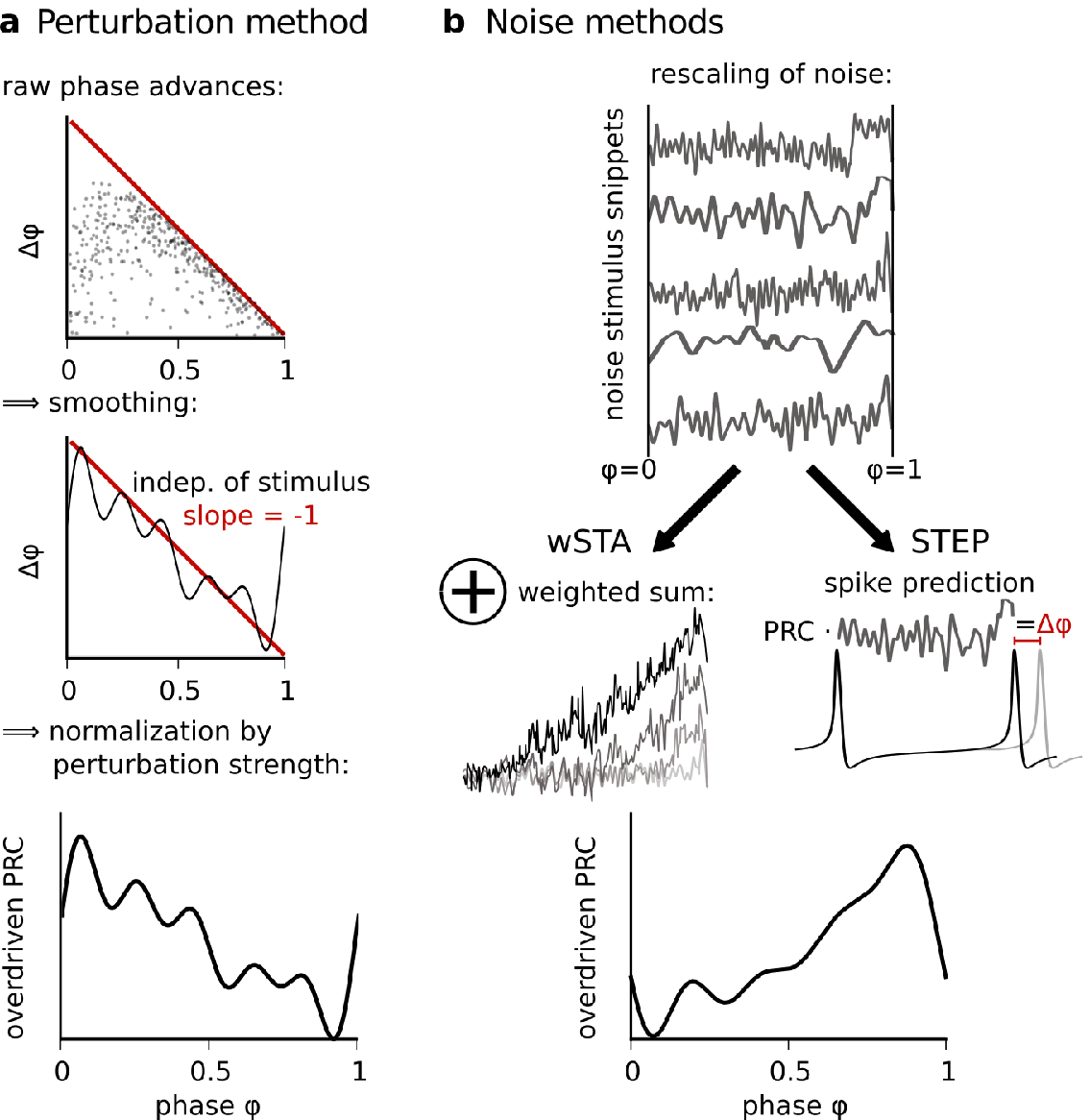}
\end{center}
\caption{Perturbation and noise-stimulation methods lead to different but
stereotypic PRCs. a: For the perturbation method, the overdriven PRC
result from the advance of the spike to the time point of the
perturbation. b: The shape of the overdriven PRC results for the noise
stimulation methods from temporally stretched noise snippets with an
elevation before the following spike.}
\label{fig:overdriven}
\end{figure*}

For the perturbation method (Figure~\ref{fig:PRCs}c), the instantaneous
initiation of spikes results in larger phase advances for earlier perturbation phases, which directly translate into a linear decrease, see
Figure~\ref{fig:overdriven}a. For the overdriven PRC in units
of phase deviation, the slope of the PRC is minus one, which has been termed the
\emph{causality limit} \citep{netoff_experimentally_2011} or \emph{causal limit} \citep{wang_hippocampal_2013}. 

For the noise methods, spikes are mostly induced by spontaneous, large deviations of the noise stimulus. As a result, the noise snippets show an exceptionally large amplitude right before the spike, i.e., close to phase one (Figure~\ref{fig:overdriven}b). Adding those
snippets up in a weighted spike-triggered average, transfers the large amplitude close to phase one to the PRC, which eventually results in a linearly increasing overdriven PRC (Figure~\ref{fig:PRCs}d). Also
for the STEP method, this temporal stretching induces a bias. As the
neuron seems to react particularly sensitively to inputs right before the
spike, it again induces a large PRC amplitude close to a phase equal to
one.

How easily an overdriven PRC can be confounded with a real PRC depends on the combination of true PRC and estimation method. For example, the overdriven PRC of the perturbation method can be easily mistaken for the PRC of a HOM spike generation, while the overdriven PRCs of the noise methods might be mistaken for the PRC of a Hopf spike generation.

\subsection{Dependence on internal
noise}\label{dependence-on-internal-noise}

When measuring PRCs in biological neurons, recordings will be perturbed
by various intrinsic noise sources including ion channel noise and recording
noise \citep{manwani_detecting_1999}. We next test the stability of PRC estimates for models implementing intrinsic noise with a strong, yet biologically realistic standard deviation, see Appendix for details. 

\begin{figure*}[h]
\hfill
\begin{center}
\includegraphics[width=0.7\textwidth]{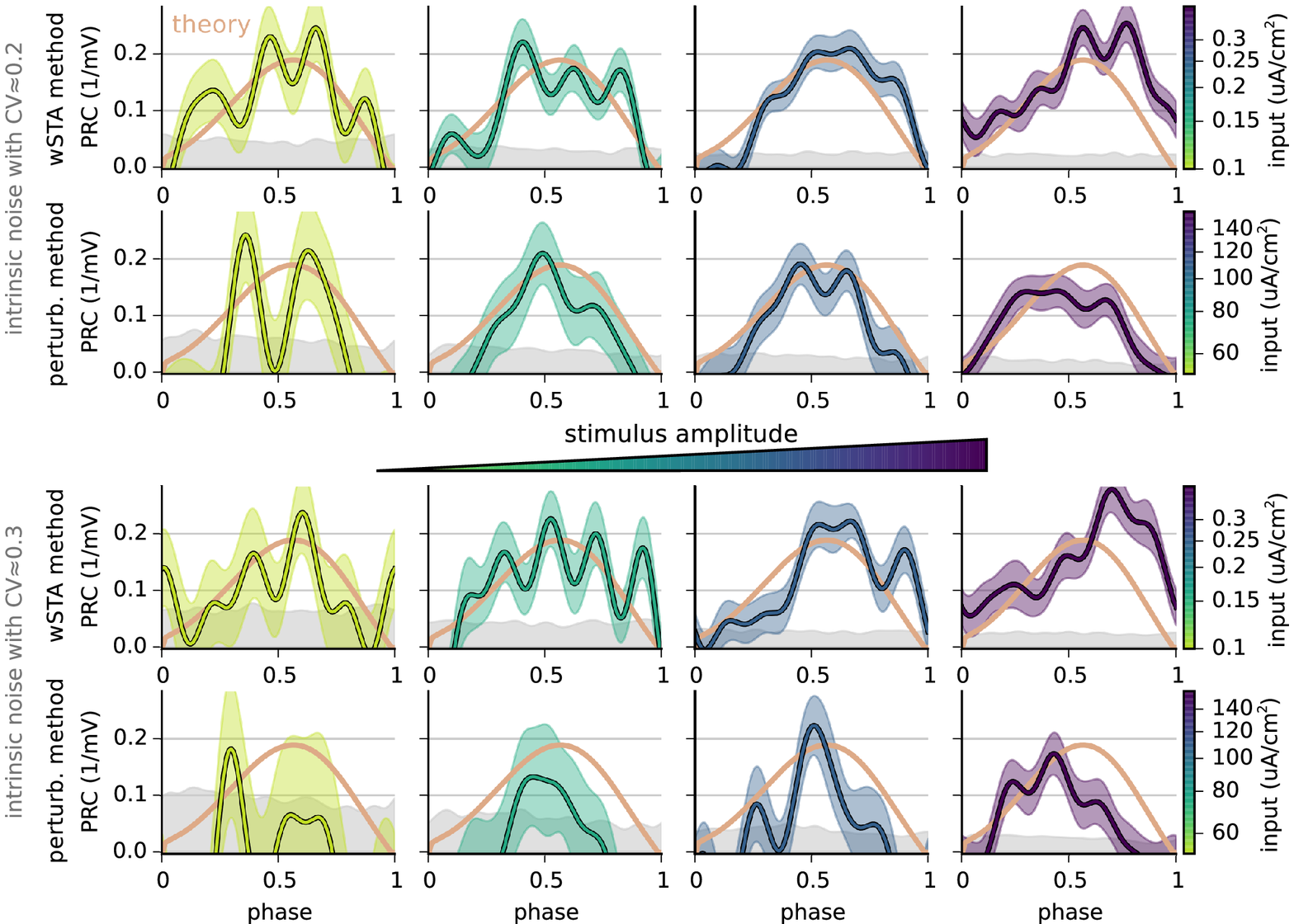}
\end{center}
\caption{Perturbation and noise-stimulation methods perform similar under high intrinsic noise levels. The PRC estimation is exemplified with the SNIC model, theoretical PRC in brown; the intrinsic noise is adapted to a baseline  coefficient of variation (CV) of 0.2 and 0.3, respectively. For the bottom traces, the PRC estimation directly reaches the overdriven regime once the stimulus amplitude is large enough to counter the intrinsic noise, not allowing for an intermediate, appropriate range of stimulus amplitudes. Bootstrapping of the data results in standard deviation errors for the PRC (transparent colored area) and for the zero-PRC background (grey area), see Appendix.}
\label{fig:noise}
\end{figure*}

With intrinsic noise, we observe an intermediate stimulus 
strength that leads to optimal PRC estimates (third column in Figure~\ref{fig:noise}). 
For small stimulus amplitudes, the stimulus' effect on spike timing is veiled by the intrinsic noise, which jitters the spikes more than the stimulus. This results in PRC estimates hardly above noise level, and the quality of the PRC estimates augments with increasing stimulus amplitude (Figure~\ref{fig:noise}, from the 
first to the third column), while the estimation error decreases.
Further increase in the stimulus amplitude leads to the
overdriven regime, first indications of which are visible in the forth
column of Figure~\ref{fig:noise}; the shift in the peak (compare Figure~\ref{fig:PRCs}a-b) will continue with increasing stimulus amplitude until the linear PRC is fully established. 
The error amplitude
continues to decrease with stimulus amplitude in the overdriven regime (data not shown). Indeed, the overdriven regime implies a stark, unrealistic reduction in both error types, which shows in published experimental results \citep{reyes_effects_1993,reyes_two_1993,goldberg_response_2007}.

A comparison between two intrinsic noise levels demonstrates that, as expected, 
larger intrinsic noise results in larger errors and increases the minimal possible stimulus
amplitude (Figure~\ref{fig:noise}, 
stronger noise in bottom panels compared to top).

Interestingly, all PRC estimation methods result in a comparable PRC quality 
(similar shape, amplitude and error levels). This result contrasts previous 
suggestions that the noise-stimulation method should be less
disturbed by intrinsic noise compared to the perturbation method
\citep{izhikevich_dynamical_2007,netoff_experimentally_2011}. In Figure~\ref{fig:noise}, even strong intrinsic noise that induces a baseline coefficient of variation (CV) of about 0.2 leads to similar PRC
estimates for all methods, while small differences might be observed for slightly higher noise
levels with a CV of about 0.3, which is close to the maximum intrinsic noise that still allows to derive PRCs.

\subsection{Practical scaling of the input signal}

When measuring PRCs from an experimentally recorded neuron, what is the correct 
amplitude of the signal? Our results suggest that the stimulus in 
addition to the DC current (i.e., perturbation or noise) should increase the firing rate by less than 10\% above baseline (i.e., when stimulated only with the DC current).
The increase in firing rate results from the positive mean value of the PRC observed for most neurons, 
which predicts that positive inputs will, on average, advance spikes to earlier
phases. The stronger the input amplitude, the larger the increase in firing rate.  In particular, our data suggests that the overdriven regime
correlates with a relative increase in firing rate exceeding 10\%.

So far, common practice recommended to aim for noise stimuli that are sufficiently
large to induce voltage deflections on
the order of 1mV, visible by bare eye \citep{netoff_experimentally_2011}. 
Yet, experimental results complying with these recommendations already show 
indications of the overdriven regime \citep{reyes_two_1993,goldberg_response_2007}.
Also in our simulations, we found that at least for
neurons with relatively high intrinsic noise, this best practice overestimates the required stimulus amplitude and thus results in overdriven PRCs. 

To bound the stimulus amplitude, one could consider the increase in spike jitter due to the stimulation, as
measured by the coefficient of variation (CV). In our numerical
simulations, the resulting CV, can
be relatively large (up to a CV of 1) without reaching the overdriven
regime, depending on the neuronal dynamics. Accordingly, we found absolute CV values to be of minor help to
establish the correct stimulus amplitude.

To summarize, a correct stimulus amplitude is indicated by
a relative increase in firing rate by maximally 10\% when adding the stimulus to the DC current. In contrast to the common assumption that the perturbing stimulus should be clearly identifiable against the intrinsic noise, the stimulus signal is in this case not visible by eye in the interspike voltage trace.

\subsection{Checking the PRC estimate}\label{preventing-overdriven-prcs}

The reporting of overdriven PRCs is undesirable, because they can severely misrepresent neuronal dynamics. To check the quality of PRC estimates, we propose  three post-experimental analyses.

As mentioned above, one hallmark of the overdriven regime is an 
unrealistically low level of estimation errors, see Sect.~\ref{dependence-on-internal-noise}. 
Thus, reported PRCs should always be complemented by meaningful error bars, like those based on bootstrapping (described in the Appendix).

The recording of PRC data for multiple stimulus amplitudes provides a good test against overdriven PRCs for all estimation methods. For the perturbation method, this was previously recommended to identify stimuli that are too weak \citep{achuthan_synaptic_2011}.
We argue that after normalization with the stimulus strength, the
amplitude of all PRC estimates with appropriate stimulus strength is
comparable, while stimuli with too large an amplitude deviate and show a larger (wSTA) or
smaller (STEP and perturbations) PRC amplitude. These size effects
are easily singled out in recordings, in contrast to the above
described alterations in PRC shape, which could be confused with the
unknown, actual PRC shape as described in Sect.~\ref{shape-of-overdriven-prcs}. 
Recording multiple stimulus amplitudes thus helps to spot overdriven PRCs, yet at the cost of a prolongation of the total recording duration. 

The noise methods provide a clear hallmark of overdriven PRCs without prolongating the recording duration when the same set of recorded spikes is evaluated with the wSTA as well as the STEP method. 
As can be seen in Figure~\ref{fig:PRCs}b, both noise-stimulus methods yield similar PRC amplitudes as long as the input strength is appropriate. Yet, the effect of overdriven stimulation on the PRC amplitude is opposite for both methods. We thus propose to estimate PRCs from the same data via STEP and wSTA
methods. A comparison of their respective amplitudes allows one to
distinguish between valid and overdriven PRC measurements: Only if the
corresponding PRCs are of similar amplitude, the measurement was
performed with an appropriate input strength. If the estimated PRCs
differ considerably in their (mean) amplitude, the measurement was most
likely overdriven. This test can also be applied to existing data, as it is based on the same experimental recording being analyzed in two different ways.

\section{Discussion}\label{conclusion}

Neuronal phase-response curves have been measured for over twenty years
\citep{reyes_effects_1993,reyes_two_1993}. PRCs are valuable to analyze neuronal dynamics, and can
be used to predict synchronization and locking behavior of the neuron. We report pitfalls of experimental PRC measurements by comparing theoretical PRCs with PRCs estimated based on simulated spike trains. We find that reliable PRC estimates require an intermediate stimulus amplitude, sufficiently high to overcome intrinsic noise, but not too high to infringe fundamental PRC assumptions. As a result of the analysis, we propose to use stimulus amplitudes that change the firing rate by not more than 10\% and that do not visibly impair the spike train. Stimulus amplitudes that are too large result in overdriven PRCs. Mild cases of overdriven PRCs are marked by a linear increase or decrease with reduced PRC error at low or high phases, respectively. Comparable hallmarks can be observed in previous experimental PRC measurements \citep{reyes_effects_1993,reyes_two_1993,goldberg_response_2007,farries_biophysical_2012,farries_phase_2012,wang_hippocampal_2013}. Even for mildly overdriven PRC estimates, information 
about neuronal dynamics may be occluded and conclusions based on these PRCs should be derived with care.

While experimental studies often refer to their estimated PRCs as ``finite'' PRCs, due to the finite, i.e., non-zero, stimulus amplitude of the experimental set-up, we here show that the distinction between ``finite'' PRC and the theoretical ``infinitesimal'' PRC is not supported by our results. We find that PRCs estimated with finite stimulus amplitude 
either fit the ``infinitesimal'' PRC, or they are not informative about neuronal dynamics: For stimuli that are too small, PRCs estimates are below noise level, and for stimuli that are too large, the PRCs approach a method-specific, generic shape that depicts the neuron misleadingly as a simplistic response machine. The transition from PRCs hardly above noise level, to PRCs with clear overdriven characteristics, is also observed in experimental studies with an increase in stimulus amplitude \citep{reyes_effects_1993,netoff_synchronization_2005}. As these experimental examples show no clear PRC estimate that lies between intrinsic-noise-dominated and overdriven, the range of appropriate, intermediate stimulus amplitudes seems to be relatively small for the recorded hippocampal and cortical cells. 

Here, we have compared PRC estimations for three methods. With optimal stimulus amplitude, all methods perform similarly \citep{torben-Nielsen_comparison_2010}, and, contrary to previous assumptions \citep{izhikevich_dynamical_2007,netoff_experimentally_2011}, we show that even under noisy conditions, the perturbation method does not \emph{per se} perform worse than the noise-stimulation methods.
Yet, the prevention of overdriven PRCs is facilitated by the noise-stimulation methods compared to the perturbation method. Noise stimulation data allows to estimate the PRC based on the wSTA as well as the STEP method, and different PRC amplitudes in both
measures provide a clear indication for overdriven PRCs. 

Overdriven PRC estimates are a common problem in experimental studies. \citet{farries_biophysical_2012,farries_phase_2012} report that even relatively large stimuli lead to reliable PRC estimates. Their solution to the problem of overdriven PRCs is a removal of the data points close to the causality limit. Others have related the causality limit to a biased sampling of phase \citep{phoka_new_2010} or phase advance data \citep{wang_hippocampal_2013}. The authors propose methods to estimate the total distributions from the available, partial observations in order to get more realistic PRC estimates. While these approaches may help to extract more information from overdriven PRCs than the traditional analysis, they do not address the alterations in phase response due to the excessive stimulus amplitude discussed here.

The PRC amplitude is rarely reported in the literature, as many studies normalize the PRC amplitude arbitrarily, e.g., to one. This practice removes important information about the PRC: Not only is the amplitude required for any quantitative description of neuronal dynamics, it also provides a most valuable tool in testing for the overdriven regime. We stress that the correct amplitude scaling is as easily extracted from PRC data as the PRC shape itself.

Preventing overdriven PRCs becomes particularly relevant when measuring neurons with relatively high levels of intrinsic noise, such as cortical or hippocampal neurons, compared to neurons with stable repetitive firing.
High levels of noise are also common when PRCs are estimated not for individual cells, but for whole brain areas. In these cases, attempted PRC measurements often suffer from exaggerated stimulus amplitudes in order to counter the intrinsic noise. In contrast, when measured with the appropriate stimulus amplitude, PRCs enrich our set of ``diagnostic'' tools to quantify neuronal dynamics.

\begin{acknowledgements}
The study was funded by the German Federal Ministry of Education and Research (Grant No. 01GQ1403) and the German Research Foundation (GRK 1589). We thank Paul Pfeiffer for valuable feedback on the manuscript.
\end{acknowledgements}

\section*{Appendix}

\subsection*{Conductance-based neuron models}

Conductance-based neuron models were based on established models from the literature. The Hopf model was first described by Morris
and Lecar; we use the version from \citet{ermentrout_mathematical_2010},
p.~50-51. The SNIC and HOM models are versions of the Wang-Buzsaki
model \citep{wang_gamma_1996}. The membrane voltage $v$ of the models follows the dynamics

\begin{eqnarray*}
\frac{\mathrm{d}v}{\mathrm{d}t} = (I_\mathrm{in} +
 g_\mathrm{L} (E_\mathrm{L} - v) + I_\mathrm{gates})/C_\mathrm{m}.
\end{eqnarray*}

The input current is the sum of a DC current, a time-dependent stimulus and an intrinsic noise $\xi$, $I_\mathrm{in} = I_\mathrm{DC} + s(t) + \sigma \xi$. For simulations without intrinsic noise, $\sigma=0$. The model parameters are summarized in Table~\ref{tab:1}. 

\begin{table}
% table caption is above the table
\caption{Parameters of the Hopf model and the
SNIC model. The HOM model is equivalent to the
SNIC model, but with $\phi = 1.5$ and $I_\mathrm{DC} = 0.166 \mu$A/cm$^2$.}
\label{tab:1}       % Give a unique label
% For LaTeX tables use
%\begin{tabular}{|l|c|c|}
\begin{tabular}{lrr}
\hline\noalign{\smallskip}
Parameter         &                Hopf            &         SNIC  \\
\noalign{\smallskip}\hline\noalign{\smallskip}
$C_\mathrm{m}$    &     $20\mu$F/cm$^2$   &        $1\mu$F/cm$^2$  \\
$E_\mathrm{L}$    &             $-60$mV   &               $-65$mV  \\
$E_\mathrm{Na}$   &             $120$mV   &                $55$mV  \\
$E_\mathrm{K}$    &             $-84$mV   &               $-90$mV  \\
$g_\mathrm{L}$    & $2\mathrm{mS/cm}^2$   & $0.1\mathrm{mS/cm}^2$  \\
$g_\mathrm{Na}$  & $4.4\mathrm{mS/cm}^2$  &   $35\mathrm{mS/cm}^2$  \\
$g_\mathrm{K}$    & $8\mathrm{mS/cm}^2$   &   $9\mathrm{mS/cm}^2$  \\
$I_\mathrm{DC}$   & $90.76 \mu$A/cm$^2$   &   $0.212 \mu$A/cm$^2$  \\
$\phi$            &              $0.04$   &                   $1$  \\
\noalign{\smallskip}\hline
\end{tabular}
\end{table}

\subsubsection*{Gating for the Hopf model:}

For the Hopf model,
\begin{eqnarray*}
I_\mathrm{gates}(v,n) = g_\mathrm{Na} m_{\infty}(v) (E_\mathrm{Na} - v) + g_\mathrm{K} n (E_\mathrm{K} - v). 
\end{eqnarray*}

The kinetics of the ion channel gating $n$ are given by
\begin{eqnarray*}
\frac{\mathrm{d}n}{\mathrm{d}t} = \phi \frac{n_\infty(v) - n}{\tau_n(v)},
\end{eqnarray*}

with $\tau_n(v) = {1\mathrm{ms}}/{\cosh \left( (v/\mathrm{mV} - 2)/60 \right) }$.
The ion channel activation curves are given as
\begin{eqnarray*}
 m_\infty(v) &=& 0.5(1+\tanh((v/\mathrm{mV} - (-1.2))/18)), \\
 n_\infty(v) &=& 0.5(1+\tanh((v/\mathrm{mV} - 2)/30)). 
\end{eqnarray*}

\subsubsection*{Gating for the SNIC and HOM models:}

For the SNIC model,
\begin{eqnarray*}
I_\mathrm{gates}(v,n,h) = g_\mathrm{Na} m_{\infty}(v)^3  h (E_\mathrm{Na} - v) + g_\mathrm{K} n^4 (E_\mathrm{K} - v). 
\end{eqnarray*}

The kinetics of the ion channel gating are given by
\begin{eqnarray*}
\frac{\mathrm{d}h}{\mathrm{d}t} &=& \phi (\alpha_{\textrm{h}}(v) (1-h) - \beta_{\textrm{h}}(v) h),\\
\frac{\mathrm{d}n}{\mathrm{d}t} &=& \phi (\alpha_{\textrm{n}}(v) (1-n) - \beta_{\textrm{n}}(v) n),
\end{eqnarray*}
with
\begin{eqnarray*}
\alpha_{\textrm{h}}(v) &=& 0.07 \exp(-\frac{v/\mathrm{mV}+58}{20})/\mathrm{ms},\\
\beta_{\textrm{h}}(v) &=& \frac{1}{ 1 + \exp(-0.1 v/\mathrm{mV}-2.8) }/\mathrm{ms},\\
\alpha_{\textrm{n}}(v) &=& \frac{0.01 v/\mathrm{mV}+0.34}{1 - \exp(-0.1 v/\mathrm{mV}-3.4)}/\mathrm{ms},\\
\beta_{\textrm{n}}(v) &=& 0.125 \exp(-\frac{v/\mathrm{mV}+44}{80})/\mathrm{ms},\\
\end{eqnarray*}

The ion channel activation curve for the gating variable $m$ is given as
\begin{eqnarray*}
 m_\infty(v) &=& \frac{\alpha_{\textrm{m}}(v)}{\alpha_{\textrm{m}}(v) + 4 \exp(-\frac{v/\mathrm{mV}+60}{18})},\\
\alpha_{\textrm{m}}(v) &=& \frac{0.1 v/\mathrm{mV}+3.5}{1 - \exp(-0.1 v/\mathrm{mV}-3.5)}.
\end{eqnarray*}

The HOM model is equivalent to the SNIC model, but with $\phi = 1.5$ and $I_\mathrm{in} = 0.166 \mu$A/cm$^2$.

Conductance-based neuron models were simulated numerically with the simulation environment \emph{brian2}
\citep{stimberg_equation-oriented_2014}. To record 500 perturbed spikes, the recording duration is 50 seconds for the noise-stimulation methods, and 100 seconds for the perturbation method, where only every second spike is perturbed. We observed that the interspike interval
depends on the time resolution $dt$ of the simulation. We have
chosen $dt$ = 0.001ms for all models, to ensure that a three
times smaller time step changed the deterministic interspike interval by less than 5\%.

To investigate the tolerance of PRC measurements towards an unknown
noise source, PRCs were in a second step measured for neuron models that included
intrinsic noise, implemented as an additive zero-mean white-noise
current (the \emph{brian2}-implemented variable \texttt{xi}). 
The noise levels chosen in this study correspond to relatively strong intrinsic noise, with a phase noise of $\tilde{\sigma} = 2\sqrt{\textrm{ms}}$ or $\tilde{\sigma} = 3\sqrt{\textrm{ms}}$ standard deviation (Figure~\ref{fig:noise}). The next section shows how to translate the phase noise into the standard deviation of the \emph{brian2}-implemented noise variable \texttt{xi}. The chosen phase noise results in a CV of around 0.2 and 0.3, respectively, which is within a biologically realistic range.

As a side-note, while the stimulus amplitude appropriate for PRC estimation is larger for perturbations than for a noise stimulus, the latter induce more spike jitter (larger CVs) compared to the perturbations.
It seems that as
the perturbation is temporally precise, even small spike deviations are
sufficient to estimate PRCs, while the continuous input delivered by the
noise stimulus requires larger deviations to estimate the PRC, probably
because every spike informs about the full phase range, instead of just
one particular phase.

\subsection*{Adaptation of intrinsic noise levels for brian2 models}\label{phaseNoise}

Comparable levels of the intrinsic noise between models can be
ensured by identical noise levels in the phase reduction. This requires
the following choice of the standard deviation of the intrinsic noise
current, $\sigma$: The deterministic phase equation Eq.~(\ref{eq:phase}) is
$\dot{\varphi}(t) = 1 + Z s(t)$ with current input $s(t)$ and PRC
$Z$. With the intrinsic noise as input, and taking a temporal mean, we
get $\dot{\varphi}(t) = 1 + \tilde{\sigma} \eta$, with the variance
of the white noise $\eta$ as
$\tilde{\sigma}^2 = \sigma^2 \int_0^1 Z^2(\varphi) \mathrm{d}\varphi$. Setting the
phase noise $\tilde{\sigma}$ to the same value in all models, we find
the appropriate current noise strength as
$\sigma = \tilde{\sigma} \frac{C_{\mathrm{m}}}{T} \left[\int_0^1 Z^2(\varphi)\right]^{-0.5} \textrm{d}\varphi$,
where $C_{\mathrm{m}}$ is the membrane capacitance of the model. For
the simulations, we evaluate this formula with the theoretical PRCs
gained from backwards integration of the adjoint equation as $Z$, see Section~\ref{backward-integration-of-the-adjoint-equation}.

\subsection*{Error estimation by
bootstrapping}\label{error-estimation-by-bootstrapping}

In order to evaluate the quality of the PRC, we use a bootstrap approach
to estimate errors for the phase response curves. Two different kinds of phase-dependent errors are considered, the baseline error which results from PRC estimates on shuffled data, and the error on the PRC estimate. 

In order to provide an
error for the PRC, we repetitively estimate PRCs using only a restricted
amount of spikes. We estimated PRCs in 100 repetitions from a set about 250 spikes, randomly chosen from the total set of about 500 perturbed spikes. The standard
deviation of the 100 PRC estimates was used as PRC error. 

In order to provide a
baseline for the PRC estimate -- above which a significant PRC should
rise -- we measure PRCs for random combinations of noise snippets and
spike advances. We measure the PRC in 100 repetitions for a random
permutation of the numbering $i$ of spike deviations $\{\Delta\varphi_i\}$, and calculate the standard
deviation for the resulting 100 estimates. The mean of these estimates
results in a PRC close to zero, such that the resulting standard
deviation corresponds to the range within which a zero, i.e., non-significant PRC estimate is to be
expected. This error around zero provides a lower bound to PRCs
that are significantly different from zero.

\subsection*{Amplitude scaling for estimated phase-response
curves}\label{correct-amplitude-scaling-for-estimated-phase-response-curves}

The appropriate unit for most experimentally derived PRCs is cm$^2$/$\mu$A, which naturally arises for PRCs that measure the phase advance in response to a current, such as a noise stimulus or perturbation in $\mu$A/cm$^2$. 
For this study, however, we aim at comparing experimentally measured PRCs with theoretical PRCs.
The theoretical PRCs gained
from backwards integration of the adjoint equation have a unit of 1/mV, as they quantify the phase
advance in response to a voltage perturbation in units of mV. For comparison, we transform the estimated PRCs into the units of the
theoretical PRCs by dividing the current PRC by the membrane
capacitance and the time window of the input,
$Z(\Delta v) = Z(\Delta I) \; C_\mathrm{m}/\tilde{\tau}$, where
$\tilde{\tau}$ is the temporal resolution of the measured PRC.
$\tilde{\tau}$ corresponds for the perturbation methods to the
duration of an individual perturbation (in our case
$\tilde{\tau}=0.1\mathrm{ms}$), and for the noise-stimulation method to
the time bin used to evaluate the PRC (i.e., spiking period divided by
the number of data points per PRC estimate, in our case
$\tilde{\tau}=T/200$).

\end{document}